\documentclass{aip-cp}

\usepackage[numbers]{natbib}
\usepackage{graphicx}
\usepackage{url}

\begin{document}

\title{Nuclear Reactions For Nucleosynthesis Beyond Fe}

\author[aff1,aff2,aff3]{Thomas Rauscher}

\affil[aff1]{Centre for Astrophysics Research, University of Hertfordshire, Hatfield AL10 9AB, United Kingdom.}
\affil[aff2]{Department of Physics, University of Basel, 4052 Basel, Switzerland.}
\affil[aff3]{UK Network for Bridging Disciplines of Galactic Chemical Evolution (BRIDGCE), \url{http://www.astro.keele.ac.uk/bridgce}, United Kingdom}

\maketitle

\begin{abstract}
Many more nuclear transitions have to be known in the determination of stellar reactivities for trans-iron nucleosynthesis than for reactions of light nuclei. This requires different theoretical and experimental approaches. Some of the issues specific for trans-iron nucleosynthesis are discussed.
\end{abstract}

\section{INTRODUCTION}
Nuclei beyond Fe are thought to have been created in a number of processes and in several different sites. The bulk of these intermediate mass and heavy nuclides were made in neutron capture processes, avoiding the problem of having to overcome high Coulomb barriers which otherwise suppress nuclear reactions in stellar plasmas. Historically, two main processes have been distinguished, the s-process (slow neutron capture) and the r-process (rapid neutron capture), each contributing about half of the abundances found beyond Fe. In the current understanding, however, these are split into subprocesses occuring in different sites \cite{advances}. For the s-process, there is a main component made in AGB stars and a weak component from massive stars \cite{kapgall}. The main component itself is made in two distinct environments inside the AGB star, during He-shell flashes and in interburst, intershell H-burning. The origin of r-process nuclei is less well understood but there are indications that also contributions from at least two different sites may be required to reproduce the r-process patterns found in the Sun and in stars of different ages \cite{arngorr,thi11}. While the long-time favored core-collapse supernovae (ccSN) were found to have problems producing r-process nuclei, outflows from neutron-star mergers and magnetically driven jets from supernova explosions have attracted recent attention because models show promise for obtaining the required conditions for an r-process. A small number ($32-35$, depending on the model used) of proton-rich nuclei cannot be made in either s- and r-process and require other production mechanisms \cite{p-review}. For most of these it has been shown that they can be made by photodisintegration of pre-existing nuclei in the outer layers of a massive star before and during its ccSN explosion. The lightest of these so-called p-nuclei, however, cannot be produced in sufficient quantities in this $\gamma$-process and call for another explanation. Various suggestions have been made, such as a $\gamma$-process in type Ia supernovae (thermonuclear disruption of a White Dwarf in a binary system), a $\nu$p-process in the deep layers of a ccSN, or a rp-process (rapid proton capture) on the surface of mass-accreting neutron stars. Finally, $^{138}$La (and perhaps partially also $^{180}$Ta) has been shown to be produced by $\nu$-induced decays, again in the outer layers of a massive star during its explosion.

Common to all these processes is that the required conditions (temperature range, matter density, neutron-to-proton ratio, duration) are constrained by observed abundances and by nuclear physics (reactivities, which are in turn determined by masses, Coulomb barriers, spectroscopy, low-energy $\gamma$- and particle strength functions). Thus, it makes sense to classify nucleosynthesis by processes, for which the appropriate conditions may be found in more than one astrophysical site. This equally applies, of course, to stellar burning stages involving light nuclei. There are several important differences for heavier nuclei, however, which also affect which nuclear properties are relevant and what experimental and theoretical approaches can be used to determine astrophysical reaction rates. Special for nucleosynthesis beyond Fe are (i) generally higher plasma temperatures, (ii) higher matter density during the astrophysical burning process, (iii) higher intrinsic nuclear level density of the involved intermediate and heavy nuclei, and (iv) higher Coulomb barriers.

Items (i) and (iv) result in higher effective interaction energies than in light element nucleosynthesis (but still low by nuclear physics standards), ranging from $8-30$ keV for the main s-process and up to 90 keV for the weak component. Neutron captures in the r-process occur at around $80-120$ keV. In charged particle reactions the effective energies are shifted to a few MeV, depending on the specific process and the charge of the nucleus and projectile (up to $\approx 9$ MeV for $\alpha$-reactions in the production of p-nuclei).
Points (i) and (ii) also imply that (except for the s-process) short-lived nuclides are involved, which can be studied only in a limited manner in the laboratory, if at all. The higher inherent nuclear level density of heavier nuclei combined with the higher temperature leads -- items (i) and (iii) -- to considerable contributions of transitions from excited states of the target nuclei (see next section).

Put in simple terms, these circumstances conspire to simplify a theoretical treatment (with exception of reactions on magic nuclei and close to the driplines) and complicate an experimental constraint of \textit{stellar} rates. The many transitions (from target states to final states, mostly via compound states) involved in explosive nucleosynthesis lend themselves to the use of averaged quantities in their prediction. On the other hand, the large number of transitions restricts the applicability of direct and indirect experimental approaches studying a few transitions, as usually applied in the study of light nuclei, even when dealing with stable nuclides. Finally, only reactions on targets in the ground state (g.s.) are accessible in the laboratory whereas heavy element nucleosynthesis not only involves highly unstable targets but also such in excited states.

In the following, some details of the above issues are briefly explained.

\section{Stellar cross sections and importance of thermally excited nuclei}
\label{sec:stellcs}

The stellar reaction rate (in reactions per time) for a reaction $a+A\rightarrow b+B$ in a plasma of temperature $T$ can be expressed as \cite{raureview,fow74,hwfz}
\begin{equation}
r^*(T) =  n_a n_A \sqrt{\frac{8}{\pi m_{aA}}} \left(kT\right)^{-3/2} \int_0^\infty \sigma_{aA\rightarrow bB}^*(E,T) E e^{-E/(kT)}\,dE =  n_a n_A \left< \sigma^* v \right>_{aA\rightarrow bB} = n_a n_A R^*(T)\quad,
\label{eq:stellrate}
\end{equation}
with the number densities $n_a$, $n_A$, the reduced mass $m_{aA}$ and Boltzmann constant $k$. The above equation also defines the reactivity $R^*=\left< \sigma^* v \right>_{aA\rightarrow bB}$ which is often used synonymously with the term ``rate''. Unfamiliarly in the nuclear physics context, the stellar cross section $\sigma^*(E,T)$ does not only depend on interaction energy but also on the plasma temperature which determines the population of excited states in the target. Explicitly showing the transitions between states $i$ in nucleus $A$ and final states $j$ in $B$, it can be written as \cite{raureview}
\begin{equation}
\sigma^*(E,T)=\frac{1}{G(T)} \sum_i \sum_j (2J_i+1) \frac{E-E_i}{E}
\sigma^{i \rightarrow j}(E-E_i) = \frac{1}{G(T)} \sum_i \sum_j W_i \sigma^{i \rightarrow j}(E-E_i)
\quad
\label{eq:stellcs}
\end{equation}
The partial cross sections $\sigma^{i \rightarrow j}$ are evaluated at the energy $E-E_i$, noting the convention by \cite{fow74} to zero cross sections at non-positive energy. Target levels are characterized by their spin $J_i$ and excitation energy $E_i$, with the g.s.\ having $i=0$ and $E_0=0$ MeV. The nuclear partition function of the target is 
\begin{equation}
G(T)=\sum_i \left( 2J_i+1\right) e^{-E_i/(kT)} = \sum_i P_i \quad.
\end{equation}

It is often overlooked that the relative importance of transitions from excited states is given by the weights $W_i=1-E_i/E$, which show a linear dependence on excitation energy, despite of the exponential decline of the thermal Boltzmann population factors $P_i$. Relying solely on the $P_i$ as indication of the importance of thermally excited states will strongly underestimate their contribution.


\begin{figure}
\includegraphics[width=.54\textwidth]{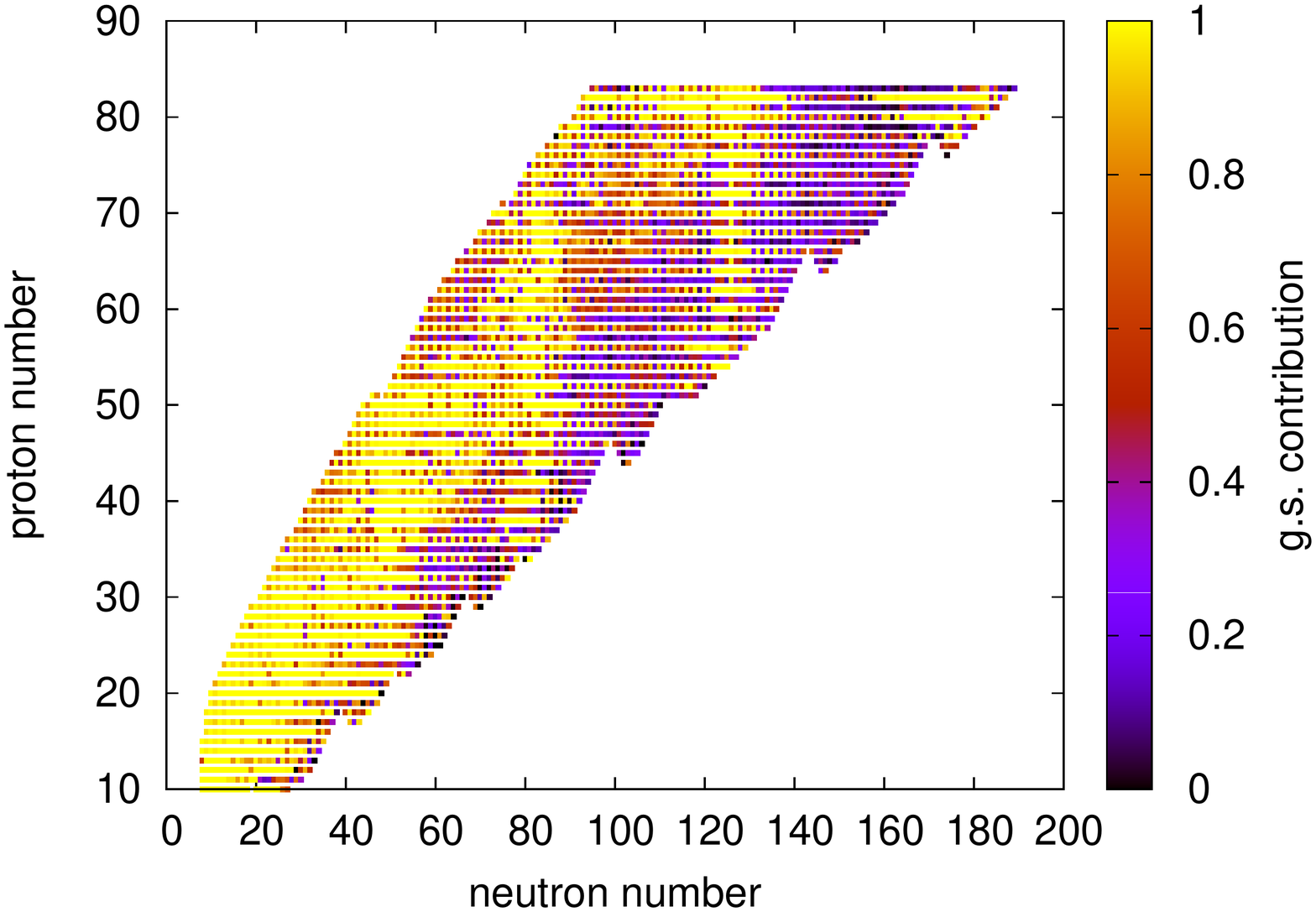}
  \hfill
\includegraphics[width=.54\textwidth]{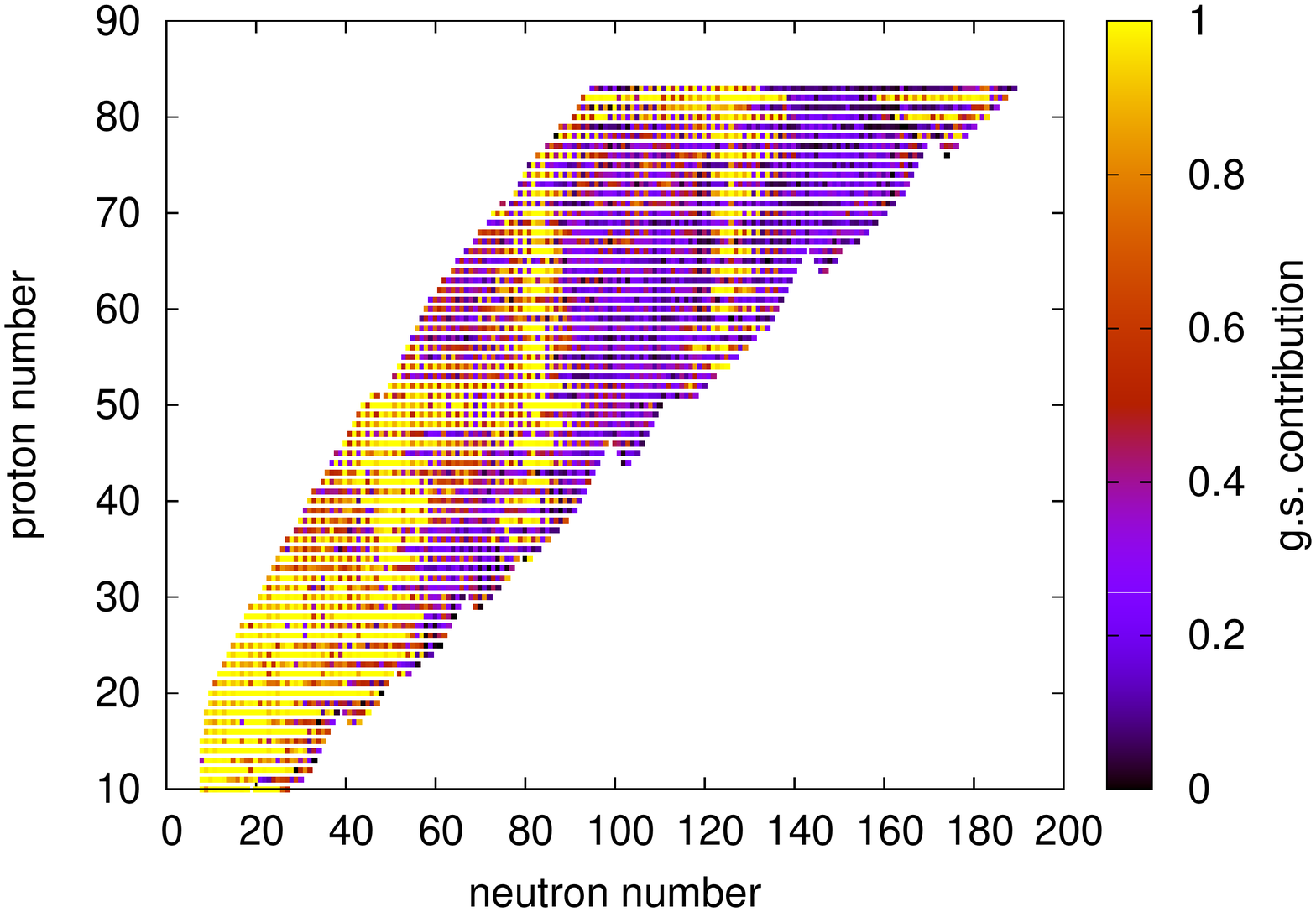}
\caption{Ground state contribution $X$ to stellar (n,$\gamma$) rates at 0.384 GK (left) and at 1 GK (right) \cite{sensis}. \label{fig:Xng}}
\end{figure}

Another instructive way to judge the importance of excited target state transitions is to look at the g.s.\ contribution to the stellar rate which is defined as \cite{Xs}
\begin{equation}
\label{eq:gsxfactor}
X_0(T)=\frac{2J_0+1}{G(T)} \frac{\int\sigma_0(E)E e^{-E/(kT)}dE}{\int\sigma^*(E,T) E e^{-E/(kT)} dE} = \frac{2J_0+1}{G(T)} \frac{R_0}{R^*} \quad,
\end{equation}
with $\sigma_0(E)=\sum_j \sigma^{0\rightarrow j}$ being the g.s.\ reaction cross section as usually determined experimentally. It always holds that $0\leq X_0\leq 1$.
It is very important to note that this is different from the simple ratio $R_0/R^*$ of g.s.\ and stellar reactivities, respectively, which has been used mistakenly in the past to quantify excited state contributions. As an example, Fig.\ \ref{fig:Xng} shows $X_0$ for neutron captures at 0.384 GK (30 keV effective energy, typical for the s-process) and 1 GK (typical for the r-process). It can be seen easily how the importance of excited state transitions increases for heavier nuclei. Even at s-process temperatures g.s.\ contributions can already be small, especially in the rare-earth region. The implications of this concerning experimentally unconstrained rate uncertainties have been discussed in \cite{Xs}. The g.s.\ contributions also decrease strongly with increasing temperature. The only exceptions are reactions close to magic numbers which retain larger g.s.\ contributions also at higher $T$ owing to the larger level spacing. Regarding charged particle reactions, $X_0$ are even smaller in most cases, due to the shift of effective interaction energy to higher energies in the integration for the rate (Eq.\ \ref{eq:stellrate}) \cite{energywindows}. All $X_0$ are given in \cite{sensis}.

As a rule of thumb when comparing g.s.\ contributions between forward reaction $a+A\rightarrow b+B$ and reverse reaction $b+B \rightarrow a+A$, the reaction direction with positive reaction $Q$-value almost always shows larger $X_0$. Among the comparatively few exceptions (due to the Coulomb suppression effect of excited state transitions \cite{coulombsupp}) are capture reactions, which always have much larger $X_0$ than their photodisintegration counterparts. In fact, $X_0<5\times 10^{-5}$ for astrophysically relevant photodisintegration rates \cite{raureview,sensis} (see also Section III.C.1 in \cite{advances} for a detailed discussion of the problematic application of photodisintegration experiments to astrophysical topics).

The $X_0$ can also be used to calculate a new uncertainty when combining experimental and theoretical information in an improved stellar rate and thus demonstrate how much a stellar rate can be constrained by a measurement at best \cite{advances}. Refs.\ \cite{sensis,Xs} also provide the recommended procedure for including measured $\sigma_0$ in an improved stellar rate. In a nutshell, instead of correcting the experimental reactivity by some factor accounting for the thermal population of excited states, the theoretical stellar reactivity $R^*$ has to be corrected by a factor $f^*$ accounting for the experimental information on the g.s.\ reactivity $R^\mathrm{exp}_0$. The new stellar reactivity $R^*_\mathrm{new}(T)=f^*(T)R^*(T)$, combining theory and experiment, is obtained by using the factor
\begin{equation}
f^*(T)=1+X_0(T)\left(\frac{R_0^\mathrm{exp}}{R_0^\mathrm{theo}}-1\right) \quad.\label{eq:renormstellar}
\end{equation}
Obviously, also the remaining uncertainty of the stellar reactivity is affected by this procedure. The new uncertainty factor $u^*_\mathrm{new}$ of $R^*_\mathrm{new}$ is constructed from the original (theory) uncertainty factor $u^*>1$ and the experimental uncertainty factor $U_\mathrm{exp}>1$ of $R^\mathrm{exp}_0$ applying
\begin{equation}
u^*_\mathrm{new}(T)=U_{\mathrm{exp}}+(u^*(T)-U_{\mathrm{exp}})(1-X_0(T)) \label{eq:uncertainty} \quad.
\end{equation}

\section{Reaction mechanisms and important nuclear properties}

Apart from increasing the contribution of excited state transitions in the stellar rate, the fact that heavier nuclei have a higher nuclear level density (LD) and thus smaller level spacing than light species has another important consequence.
The LD at the formation energy $E_\mathrm{form}$ of the compound nucleus $a+A=C=b+B$ determines the dominant reaction mechanism. In the absence of levels to be populated close to the given energy, \textit{direct reactions} to low-lying final states dominate the reaction cross section \cite{raureview,MaMe83,drc}. When there is a small number of well separated excited states close to $E_\mathrm{form}$ in $C$, this gives rise to resonances in the reaction cross sections, which can be described by the Breit-Wigner formula or by the R-matrix method \cite{raureview,lt58}. The considerable challenge therein is to determine the properties of the resonances contributing to the reaction rate integral. It is experimentally challenging to perform the required measurements for unstable nuclei and/or at low energy and theoretical \textit{ab initio} methods cannot be applied to heavy nuclei, yet.

Finally, an extreme case of resonant reactions appears when the LD is high, leading to a large number of unresolved resonances which can be described by averaged resonance parameters. This is called the Hauser-Feshbach approach \cite{haufesh,adndt}. The vast majority of reactions in nucleosynthesis of intermediate and heavy nuclei can be described in this model \cite{raureview,rtk} Despite of the increased number of nuclear transitions to be included in the model, this facilitates the predictions somewhat, as averaged quantities can be used. Calculations of reaction rates from smooth Hauser-Feshbach cross sections are also somewhat more ``forgiving'' to fluctuations around the ``true'' cross section value because of the integration over the projectile energy distribution. Different quantities, though, have to be known as input to the calculations than for direct or resonant reactions, such as optical potentials (related to the effective interaction in a many-nucleon system) and level densities (but not, e.g., spectroscopic factors of isolated levels) \cite{raureview}.
Nevertheless, Breit-Wigner and Hauser-Feshbach cross sections are closely related and the expressions for the cross sections look very similar \cite{raureview}. Also, the reactivities depend on similar expressions in both cases, a fraction
\begin{equation}
\label{eq:f}
\mathcal{F}(E_C,J_C,\pi_C)=(2J_C+1)\frac{W_{E_C,J_C,\pi_C}^A W_{E_C,J_C,\pi_C}^B}{W_{E_C,J_C,\pi_C}^\mathrm{all}} \quad.
\end{equation}
The quantities $W^A$, $W^B$, and $W^\mathrm{all}$ are the total width of a compound level in $C$ with spin $J_C$ at excitation energy $E_C$ and including all transitions to levels $i$ in nucleus $A$ with $E_i \leq E_C-S_\mathrm{pro}$ in the entrance channel, the total width of this compound state including all possible transitions to levels $j$ in $B$ with $E_j \leq E_C-S_\mathrm{ej}$ in the exit channel, and the combined width $W_{E_C,J_C,\pi_C}^\mathrm{all}=\sum_{\lambda=A, B, \dots} W_{E_C,J_C,\pi_C}^\lambda$ from the sum of all emission processes from the compound state to all channels $\lambda$, respectively \citep{hwfz,raureview}. The separation energies of the projectile and the ejectile in the compound nucleus $C$ are denoted by $S_\mathrm{pro}$ and $S_\mathrm{ej}$, respectively. Assuming narrow resonances, the rate of a single Breit-Wigner resonance at excitation energy $E_C$ (this would be at an energy of $E=E_C-S_\mathrm{pro}$ in the entrance channel) is given by \citep{raureview,sensis}
\begin{equation}
\label{eq:BWrate}
r_\mathrm{BW}^* \propto \frac{2J_0+1}{G}\frac{1}{(kT)^{3/2}} e^{\left( E_C-S_\mathrm{pro} \right) /(kT)} \mathcal{F}(E_C,J_C,\pi_C) \quad.
\end{equation}
In the Hauser-Feshbach model it is assumed that there is a large number of resonances at any compound formation energy $E_C$ and for any $J_C$ and parity $\pi_C$. Therefore, the rate is
\begin{equation}
\label{eq:HFrate}
r_\mathrm{HF}^* \propto \frac{2J_0+1}{G}\frac{1}{(kT)^{3/2}} \int_0^\infty \sum_{J_C,\pi_C} \mathcal{F}(E+S_\mathrm{pro},J_C,\pi_C) E  e^{-E/(kT)}\,dE\quad.
\end{equation}
In this case, the widths used in $\mathcal{F}$ are assumed to be averaged quantities for a large number of similar resonances.

The similarity of $\mathcal{F}$ as defined in Eq.\ (\ref{eq:f}) in the Breit-Wigner and Hauser-Feshbach approaches leads to the expectation of similar dependences on input quantities, i.e., on (averaged) widths. This is actually the case and allows to better disentangle the seemingly complicated sensitivity of Hauser-Feshbach predictions on input quantities such as masses, optical potentials, $\gamma$-strength functions, and LDs. Often, these dependences are explored by arbitrarily using a selection of predefined treatments of these quantities and observing the change in the resulting cross sections or reactivities. This is then taken as an indication of the uncertainty in the underlying model. From a theory point of view, this is an inadequate or even incorrect procedure as an arbitrary selection of input models can never catch the actual uncertainty (see \cite{sensis} for error estimates in predictions). An uneducated use of models may be misleading also because some of the models may have known limitations, discouraging the application to the chosen reaction. Rather, a \textit{systematic} variation of the fundamental quantities, i.e., the reaction widths, is called for. Similarly, when trying to model a measured cross section, the starting point should be systematic variations of widths. After having identified the actually relevant widths and their required changes, it can be discussed how to obtain these based on theory.

\begin{figure}
\includegraphics[width=0.7\columnwidth]{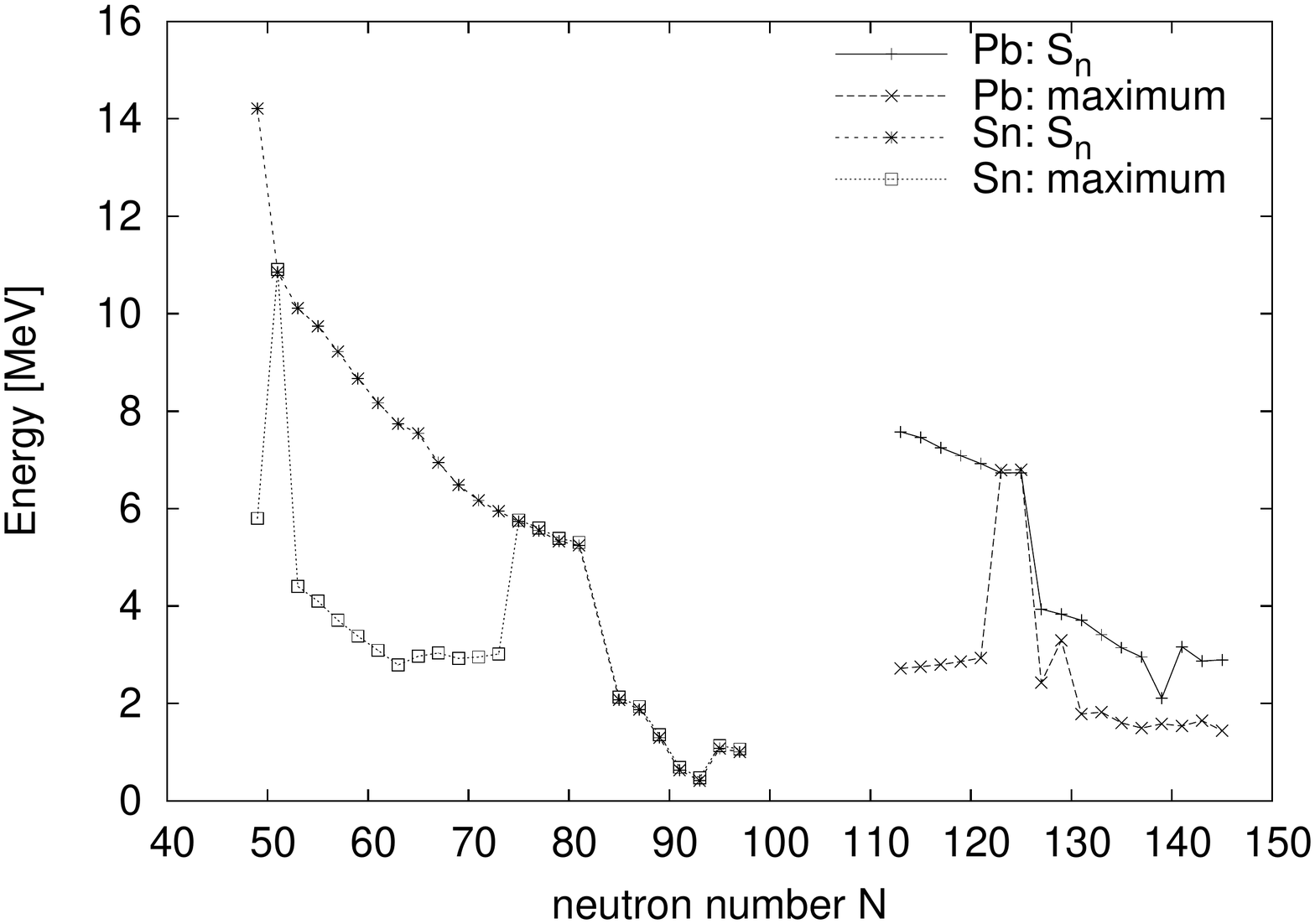}
\caption{\label{fig:snpb_n_even}The maximally contributing E1 $\gamma$ energies when capturing 60 keV neutrons on Sn and Pb isotopes with even mass numbers are compared
to the neutron separation energies $S_\mathrm{n}$ in the compound nuclei \cite{raugamma}. Note that the horizontal axis gives the neutron number $N$ of the compound (final) nucleus.}
\end{figure}

For \textit{systematic} sensitivity studies it is convenient to define the sensitivity \cite{sensis}
\begin{equation}
\label{eq:sensi}
s=\frac{v_\Omega-1}{v_q-1} \quad.
\end{equation}
It is a measure of a change by a factor of $v_\Omega=\Omega_\mathrm{new}/\Omega_\mathrm{old}$ in the output $\Omega$ as the result of a change in the input quantity $q$ by the factor $v_q=q_\mathrm{new}/q_\mathrm{old}$, with $s=0$ when no change occurs and $s=1$ when the final result changes by the same factor as used in the variation of $q$, i.e., $s=1$ implies $v_\Omega=v_q$. Negative $s$ are usually taken to indicate that $q$ and $\Omega$ vary in opposite directions. A large scale study of sensitivities for reactions with light projectiles from proton- to neutron-dripline has been published in \cite{sensis}, where g.s.\ cross sections and stellar reactivities have been chosen as $\Omega$ and particle- and $\gamma$-widths $W$, as well as LDs, were varied as $q$. While it may not be surprising that neutron captures are mainly sensitive to variations in the $\gamma$-widths, due to the $\gamma$-width being much smaller than the neutron width, this does not apply to reactions with charged particles. In stellar reactivities, charged particle widths are always smaller than both neutron- and $\gamma$-widths, due to the small interaction energies and high Coulomb barriers. This poses a big challenge for experiments as it is very difficult to impossible to determine charged particle widths at astrophysically relevant energies. Another finding was that only those reactivities are sensitive to changes in the LD which are also sensitive to the $\gamma$-width. This is because transitions to low-lying, discrete excited states dominate the particle widths (due to their larger relative energies) whereas $\gamma$-widths are dominated by transitions with smaller relative energy to levels with higher excitation energies \cite{raureview}. The latter can be understood by realizing that the competition between increasing LD with increasing excitation energy in $C$ and decreasing $\gamma$-strength (because of decreasing relative $\gamma$-energy in $C$) leads to a peak in the contribution of single $\gamma$-transitions at a specific energy $E_\gamma^\mathrm{peak}$. Interestingly, $E_\gamma^\mathrm{peak}\approx 2-4$ MeV for astrophysical neutron- as well as charged particle capture \cite{raugamma}. This rule is only broken for reactions with low $E_\mathrm{form}$ in nuclei with low LD, such as at magic numbers or close to driplines, where $\gamma$-transitions to the g.s.\ dominate. Figure \ref{fig:snpb_n_even} shows an example for neutron captures on even Sn and Pb isotopes. More details are found in \cite{raugamma}. The knowledge of the contributing $\gamma$-energies is important for judging the relevance of modifications to the photon-strength function (such as pygmy resonances) and the LD. Changes will affect reactivities only when they appear at energies around $E_\gamma^\mathrm{peak}$, otherwise they will be inconsequential for astrophysical applications.

Exceptions to the applicability of the Hauser-Feshbach model are reactions with very low or negative projectile separation energy in $C$ because then $E_\mathrm{form}$ is shifted to very low excitation energies, at which the LD may be too low to apply the model even for heavier nuclei. Typically this occurs close to the driplines. Moreover, due to the lower LD at magic nucleon numbers, the Hauser-Feshbach model also may not be applicable to magic target nuclei and low plasma temperatures \cite{rtk}. This applies mainly for neutron captures, however, because the projectile energy range relevant for the calculation of the reaction rate is shifted to higher compound excitation energies when there are charged particles in entrance or exit channel \cite{energywindows}.

The importance of direct reactions -- and specifically of direct neutron capture -- in astrophysical environments has been discussed in literature (see, e.g., \cite{raureview,MaMe83,drc} and references therein). Their sensitivities, however, are comparatively simple to estimate because only single transitions are involved instead of summed widths. Thus, the rate of a direct reaction will linearly depend on the spectroscopic factors used. They will also be more strongly sensitive to a change in the properties of the single final state (excitation energy, spin, parity, and whether it is bound or unbound) than compound reactions \citep{drc}.
Direct reactions at high energies -- (d,p) and (d,n) reactions at several tens to hundreds of MeV, for instance -- have also been suggested to be used in nuclear spectroscopy studies of unstable nuclei at radioactive ion beam facilities. Cross sections of reactions at such high energies do not appear in the calculation of astrophysical reaction rates but the extracted information on low-lying excited states and spectroscopic factors is useful in the calculation of the widths $W$ implicitly appearing in Equations (\ref{eq:BWrate}) and (\ref{eq:HFrate}).

\section{Conclusion}

Many more transitions have to be known to determine stellar reactivities for trans-iron nucleosynthesis than in light element nucleosynthesis. This requires different theoretical and experimental approaches. Due to pronounced thermal plasma effects in the stellar rates, most measurements can only support predictions by testing models for single transitions but not constrain a stellar rate independently. It has to be made sure that astrophysically relevant transitions are studied, though. Published systematic sensitivity studies and g.s.\ contributions to the stellar rate help to guide experiments. On the other hand, predictions are simplified by being able to average over many transitions and apply the Hauser-Feshbach model, which has been successful in describing a large number of reaction cross sections. The challenges for theory are twofold. First, nuclear structure models have to be improved to be able to reliably predict the nuclear properties (such as nuclear spectroscopy, LD, or $\gamma$-strength functions) required for the astrophysical reaction cross section prediction. Another major challenge to theory is to accurately describe the competition between direct, resonant, and Hauser-Feshbach reaction mechanisms for nuclei close to magic numbers and close to the driplines. Some first attempts have been made to combine direct and Hauser-Feshbach cross sections for neutron-rich nuclei \cite{raureview,xu} but currently a predictive treatment of individual resonances (which nevertheless may be very important also for magic nuclei and close to driplines) is beyond the reach of theory.

\section{ACKNOWLEDGMENTS}
This work was partially supported by the European Research Council (grant GA 321263-FISH) and the UK Science and Technology Facilities Council (grant ST/M000958/1).



\begin{thebibliography}{99}
\bibitem{advances} T. Rauscher, AIP Advances \textbf{4}, 041012 (2014). 
\bibitem{kapgall} F. K\"appeler, R. Gallino, S. Bisterzo, and W. Aoki, Rev.\ Mod.\ Phys.\ \textbf{83}, 157 (2011).
\bibitem{arngorr} M. Arnould, S. Goriely, and K. Takahashi, Phys.\ Rep.\ \textbf{450}, 97 (2007).
\bibitem{thi11} F.-K. Thielemann, et al., Prog.\ Part.\ Nucl.\ Phys.\ \textbf{66}, 346 (2011). 
\bibitem{p-review} T. Rauscher, et al.,
Rep.\ Prog.\ Phys.\ \textbf{76}, 066201 (2013).
\bibitem{raureview} T. Rauscher, Int.\ J. Mod.\ Phys.\ E \textbf{20}, 1071 (2011).
\bibitem{fow74} W. A. Fowler, Quart.\ J. Roy.\ Astron.\ Soc.\ \textbf{15}, 82 (1974).
\bibitem{hwfz} J. A. Holmes, et al.,
At.\ Data Nucl.\ Data Tables \textbf{18}, 305 (1976).
\bibitem{sensis} T. Rauscher, Astrophys.\ J. Suppl.\ \textbf{201}, 26 (2012). 
\bibitem{Xs} T. Rauscher, Astrophys.\ J. Lett.\ \textbf{755}, L10 (2012). 
\bibitem{energywindows} T. Rauscher, Phys.\ Rev.\ C \textbf{81}, 045807 (2010).
\bibitem{coulombsupp} T. Rauscher, Phys.\ Rev.\ C \textbf{88}, 035803 (2013). 
\bibitem{raugamma} T. Rauscher, Phys.\ Rev.\ C \textbf{78}, 032801(R) (2008).

\bibitem{MaMe83} G. J. Mathews, A. Mengoni, F.-K. Thielemann and W. A. Fowler, \textit{Astrophys.\ J.} \textbf{270}, 740 (1983).

\bibitem{drc} T. Rauscher, et al.,
Phys.\ Rev.\ C \textbf{57}, 2031 (1998).

\bibitem{lt58} A. M. Lane and R. G. Thomas, Rev.\ Mod.\ Phys.\ \textbf{30}, 257 (1958).


\bibitem{haufesh} W. Hauser and H. Feshbach, Phys.\ Rev.\ \textbf{87}, 366 (1952).

\bibitem{adndt} T. Rauscher and F.-K. Thielemann, At.\ Data Nucl.\ Data Tables \textbf{75}, 1 (2000).

\bibitem{rtk} T. Rauscher, F.-K. Thielemann, and K.-L. Kratz, Phys.\ Rev.\ C \textbf{56}, 1613 (1997).

\bibitem{xu} Y. Xu, S. Goriely, A. J. Koning, and S. Hilaire, Phys.\ Rev.\ C \textbf{90}, 024604 (2014).

\end{thebibliography}

\end{document}